\documentclass[floatfix,showpacs,pre,aps,a4paper,twocolumn]{revtex4}

 \usepackage{bm}
 \usepackage{amsmath}
 \usepackage{amssymb}
 \usepackage{dsfont}
 \usepackage{latexsym}
 \usepackage{amsfonts}
 \usepackage{epsfig}
 \usepackage{subfigure}
 \usepackage{color}
 \definecolor{darkblue}{rgb}{0,0,.5}
 \usepackage[linktocpage, colorlinks=true, linkcolor=darkblue, citecolor=darkblue]{hyperref}
 \usepackage[all]{hypcap}

\newcommand{\C}[1]{{\cal{#1}}}
\newcommand{\bb}[1]{\textbf{#1}}
\newcommand{\ua}[0]{{\uparrow}}
\newcommand{\da}[0]{{\downarrow}}
\newcommand{\ra}[0]{{\rightarrow}}

\begin{document}

\title{The Second Laws for an Information driven Current through a Spin Valve}

\author{Philipp Strasberg$^1$}
\author{Gernot Schaller$^1$}
\author{Tobias Brandes$^1$}
\author{Christopher Jarzynski$^2$}
\affiliation{
$^1$ Institut f\"ur Theoretische Physik, Technische Universit\"at Berlin, Hardenbergstr. 36, D-10623 Berlin, Germany\\
$^2$ Department of Chemistry and Biochemistry and Institute for Physical Science and Technology, University of Maryland, College Park, Maryland 20742, USA}

\begin{abstract}
We propose a physically realizable Maxwell's demon device 
 using a spin valve interacting unitarily for a short time with electrons placed on a tape of quantum dots, 
 which is thermodynamically equivalent to the device 
 introduced by Mandal and Jarzynski [PNAS \bb{109}, 11641 (2012)]. The model is exactly solvable 
 and we show that it can be equivalently interpreted as a Brownian ratchet demon. We then consider a measurement based 
 discrete feedback scheme, which produces identical system dynamics, but possesses a different second law inequality. We show that 
 the second law for discrete feedback control can provide a smaller, equal or larger 
 bound on the maximum extractable work as compared to the second law involving the tape of bits. Finally, we derive an 
 effective master equation governing the system evolution for Poisson distributed bits on the tape (or measurement times 
 respectively) and we show that its associated entropy production rate contains the same physical statement as the second 
 law involving the tape of bits. 
\end{abstract}


\maketitle

\section{Introduction}

Traditionally, thermodynamics is a theory that describes systems exchanging energy and entropy with idealized heat reservoirs and work 
sources. It provides two fundamental laws every system must obey: 
the first law (energy balance) and the second law (entropy increase). Reconciling these macroscopic laws
with a microscopic picture of atoms and molecules in motion raises subtle issues.
As far back as the nineteenth century, Maxwell speculated that an external agent, using microscopic information
to control the state of a system, might be able to circumvent the second law \cite{Maxwell1871}.
This agent is now known as `Maxwell's demon'. Accepting the possibility of `intelligent interventions' 
(due to idealized measurement and information processing devices such as computers), it is widely accepted that the correct application 
of Landauer's principle \cite{Landauer1961} will save the second law, as noted by Bennett \cite{Bennett1982} 
in his analysis of Szilard's famous engine \cite{Szilard1929}. 
Note however that Maxwell's demon can also be `exorcized' without reference to intelligent interventions, by showing that 
any device designed to exploit molecular scale fluctuations to violate the second law, must fail or must produce 
a large dissipation somewhere else as analysed for specific models by Smulochowski \cite{Smulochowski1912} and Feynman \cite{FeynmanLectures1963}. 
See also Ref. \cite{NortonEntropy2013} for opposing arguments and Refs. \cite{SkordosZurek1992,StrasbergSchallerPRL2013} for other 
illustrative models.

Besides the possibility of actively measuring and controlling the system, Bennett noted that a tape of bits in a low 
entropy state might act as a thermodynamic resource while the tape randomizes itself \cite{Bennett1982}, which can be regarded as another 
form of a Maxwell demon in which the low entropy state of the tape allows for the rectification of thermal fluctuations, for instance to lift a mass 
or cool a cold reservoir. A particular mathematical model was proposed and analyzed by Mandal and Jarzynski \cite{MandalJarzynskiPNAS2012}
and has been followed by further models exploring the possibility of Maxwell's demon devices incorporating an information 
reservoir in the thermodynamic description 
\cite{BaratoSeifertEPL2013, HorowitzSagawaParrondoPRL2013, MandalQuanPRL2013, DeffnerJarzynskiPRX2013, DeffnerPRE2013, HoppenauEngelEPL2014}. 

However, the models proposed above are rather abstract and 
it would be desirable to find a \emph{physical} model able to reproduce the same results. By `physical' model we mean 
that we start with a well-defined and well-motivated Hamiltonian describing an explicit, physically realizable system, and, following
standard procedures (e.g. as used to derive master equations), we end up with a mathematical description equivalent to those above.
In this paper we will indeed show that a quantum dot spin valve 
with perfectly polarized leads allowing only for one sort of spins to tunnel through, which interacts with a tape of electrons 
causing spin flips, provides a physical implementation of the device proposed in Ref. \cite{MandalJarzynskiPNAS2012} in the sense 
that it has identical dynamics and thermodynamics (though a quite different physical interpretation). A similar model was already put forward by 
Datta in Ref. \cite{DattaNotes2007} where the `impurities' in his model correspond to the electrons on the tape in our model. 
For another physical realization also see Ref. \cite{LuMandalJarzynskiPT2014}. 

The models discussed above all rely on measurements or interactions with a bit at predetermined discrete times 
or intervals. The resulting dynamics can in general not be formulated as a differential equation anymore. 
An alternative approach to feedback control relies on a continuous measurement scheme and the fact that a master equation (ME) 
can be unraveled in terms of trajectories representing the actual state of the system. This approach yields an effective differential 
equation for the system dynamics and has been extensively used in the field of 
open quantum systems and quantum information \cite{WisemanMilburnBook}. Furthermore, it was successfully applied to construct 
Maxwell demon like feedbacks and to study the thermodynamics of such systems 
\cite{SchallerEtAlPRB2011, EspositoSchallerEPL2012, StrasbergSchallerPRE2013}. It was shown that 
such feedback schemes can also be formulated within an inclusive approach not relying on any phenomenological 
measurements or feedback actions \cite{StrasbergSchallerPRL2013}. 
In addition to the ME picture, the thermodynamic implications of continuous feedback schemes were studied for Langevin dynamics as well 
\cite{MunakataRosinbergJSM2012, MunakataRosinbergJSM2013}. 

Aside from the quest to design illustrative devices, which are able to rectify thermal fluctuations by 
some sort of information processing, the question naturally arises as to whether it is possible to treat the different 
approaches above in a unified framework 
\cite{TasakiArXiv2013, BaratoSeifertPRL2014, HorowitzEspositoPRX2014, ShiraishiSagawaArXiv2014, BauerBaratoSeifertArXiv2014}. 
To address this question with our model -- following the three cases investigated by Barato and Seifert \cite{BaratoSeifertPRL2014} -- 
we will design a measurement-based feedback scheme, which has identical system dynamics as the tape model, and we will 
derive an effective ME for Poisson distributed bits on the tape or measurement times respectively. We will compare the 
different forms of the second laws and we will show that the second law for discrete feedback control can provide 
a smaller, equal or larger bound on the amount of extractable work as compared to the second law involving the tape of 
bits. For the effective ME we will show that its associated entropy production represents the same physical statement 
as the second law involving the tape of bits. 

\emph{Outline: } We start with the description of the model including the system, the tape of bits and the system-bit interaction 
in Sec. \ref{sec physical Implementation of the model} and we show that it is equivalent to a Brownian ratchet demon. 
Sec. \ref{sec thermodynamics} is then devoted to a thorough study of its thermodynamics. In Sec. 
\ref{sec measurement based feedback scheme} we will study a measurement based feedback scheme, which yields identical 
system dynamics but different thermodynamics and we will compare the two different second laws. Finally, we derive an 
effective ME for a Poisson distributed tape in Sec. \ref{sec effective master equation} and discuss its thermodynamic behaviour 
in relation to the other approaches. In Sec. \ref{sec conclusions} we discuss our results.

\section{Physical Implementation of the Model}
\label{sec physical Implementation of the model}

\subsection{System Description}

The system we want to control -- either by a tape of bits or by feedback control 
(see Sec. \ref{sec measurement based feedback scheme}) -- consists of a quantum dot in the ultra strong Coulomb blockade 
regime attached to two spin-polarized ferromagnetic leads, which preferably allow one spin direction to tunnel. 
Such quantum dot spin valves can be experimentally realized by molecular quantum dots 
\cite{BoganiWernsdorferNatMat2008, UrdampilettaEtAlNatMat2011}. 
The quantum dot is either empty or filled by an electron with spin up or spin down, thus the system Hilbert space is 
three dimensional, $\C H_S = \mbox{span}\{|0\rangle,|\ua\rangle,|\da\rangle\}$, and the Hamiltonian of the system is assumed to be 
$H_S = \epsilon_s(|\ua\rangle\langle\ua| + |\da\rangle\langle\da|)$. For perfectly and oppositely polarized leads the 
usual Born-Markov secular ME predicts that the time evolution of the probability vector 
$\rho_S = (p_0,p_\ua,p_\da)$ is given by the rate equation $\partial_t\rho_S(t) = \C W\rho_S(t)$ where the Liouvillian
$\C W$ can be split into two parts $\C W = \C W_L + \C W_R$: 
\begin{equation}\label{eq Liouvillian spin valve}
 \begin{split}
  \C W_L &= \Gamma\left(\begin{array}{ccc}
		     -f_L	&	1-f_L		&	0	\\
		     f_L	&	-(1-f_L)	&	0	\\
		     0		&	0		&	0	\\
                    \end{array}\right),	\\
 \C W_R &= \Gamma\left(\begin{array}{ccc}
		     -f_R	&	0	&	1-f_R		\\
		     0		&	0	&	0		\\
		     f_R	&	0	&	-(1-f_R)	\\
                    \end{array}\right).
 \end{split}
\end{equation}
A detailed derivation of this equation even for arbitrary polarized leads is given in Ref. \cite{BraunKoenigMartinekPRB2004} 
following the steps stated in \footnote{To obtain Eq. (\ref{eq Liouvillian spin valve}) from Ref. \cite{BraunKoenigMartinekPRB2004} 
one needs to take the limits (in their notation) of (i) ultra strong Coulomb blockade ($U\rightarrow\infty$), (ii) oppositely 
polarized leads ($\hat{\bb n}_L = \bb e_x = -\hat{\bb n}_R$), (iii) fully polarized leads ($p_L=p_R=1$) and (iv) for simplicity 
equal tunneling rates $\Gamma_L = \Gamma_R \equiv \Gamma/2$. }. 
Furthermore, within the framework of the Born-Markov secular ME, we do not have to include 
the off-diagonal elements of the system density matrix (``coherences'') because in the energy eigenbasis 
they do not affect the populations but simply follow damped oscillations and die out in the long-time limit. 
The transition rates of the Liouvillian $\C W$ are specified by a global tunneling rate $\Gamma>0$ and Fermi functions 
$f_\nu = f_\nu(\epsilon_s) = [e^{\beta_\nu(\epsilon_s-\mu_\nu)}+1]^{-1}$, $\nu\in\{L,R\}$, 
where $\beta_\nu$ and $\mu_\nu$ are the inverse temperature and chemical potential of lead $\nu$. For the rest of the text we 
fix the chemical potentials at $\mu_L \equiv \epsilon_s+\frac{V}{2}$ and $\mu_R \equiv \epsilon_s-\frac{V}{2}$ 
where $V=\mu_L-\mu_R$ denotes the bias voltage. Assuming equal lead temperatures, $\beta_L = \beta_R \equiv \beta$, 
the Fermi functions fulfill the local detailed balance condition 
\begin{equation}\label{eq local detailed balance}
 \frac{f_L}{1-f_L} = e^{\beta V/2}, ~~~ \frac{f_R}{1-f_R} = e^{-\beta V/2}
\end{equation}
and the steady state $\bar\rho_S$ of the system is given by 
\begin{equation}\label{eq steady state spin valve}
 \bar p_\ua = \frac{e^{\beta V}}{Z}, ~~~ \bar p_0 = \frac{e^{\beta V/2}}{Z}, ~~~ \bar p_\da = \frac{1}{Z}
\end{equation}
with $Z = e^{\beta V} + e^{\beta V/2} + 1$. 
It is easy to verify that the electric current through the system vanishes: due to the opposite polarization 
of the leads the transport through the spin valve is blocked. We note however that this is an idealized description. 
In practice the leads are not perfectly polarized and phonons in the surrounding medium can induce spin flips 
in the quantum dot. For simplicity we will neglect these effects because they do not change the general point of 
our discussion as long as they are relatively weak. For the rest of this paper we will set $\beta\equiv1$. 

\subsection{Tape of Bits}

We now describe the tape of bits, which moves frictionlessly past the quantum dot spin valve with constant speed. 
In contrast to the models of Refs. \cite{MandalJarzynskiPNAS2012, BaratoSeifertEPL2013, MandalQuanPRL2013, DeffnerPRE2013} 
we assume that each bit interacts with the system only for a short time $\delta t \ll \Gamma^{-1}$ and that the time 
$\tau$ between two successive bits is relatively long, $\tau\gg\delta t$. During this time $\tau$ the system 
evolves according to the Liouvillian $\C W$. 

\begin{figure}
\includegraphics[width=0.44\textwidth,clip=true]{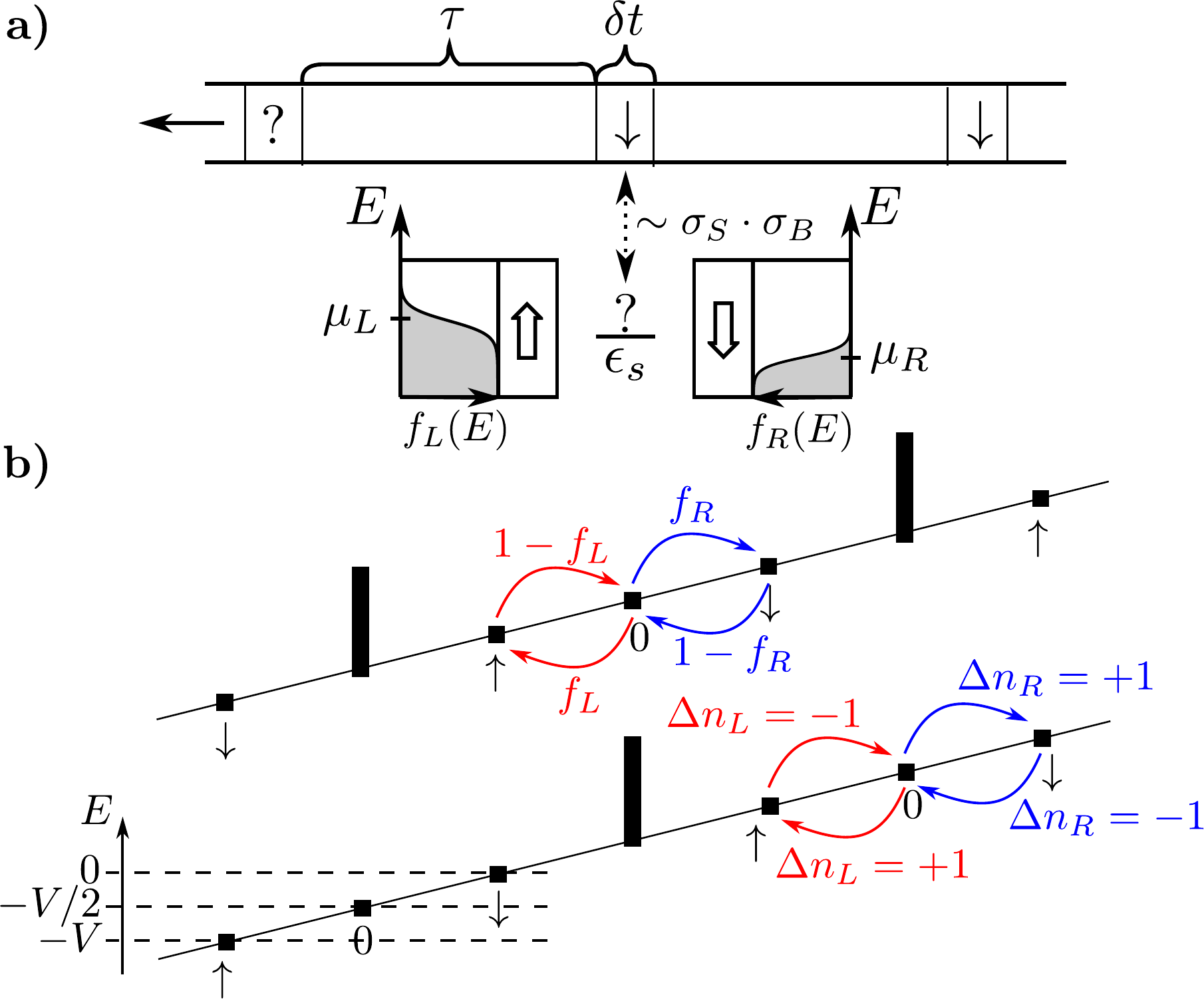}
\caption{\label{fig sketches sys ratchet} a) Simple sketch of the system showing the tape of bits coupled to the quantum dot 
spin valve. b) Equivalent model of the Brownian ratchet showing the two different modes, which can be regarded as shifted 
versions of an infinite lattice. Transitions between the two different modes are due to the interaction with a bit or due 
to the feedback. }
\end{figure}

Physically, each bit on the tape represents another quantum dot 
containing one excess electron, which is either in state spin up or spin down. To quantify the portion of 
spin up and spin down electrons we introduce the excess parameter $\delta^-$, which is defined by 
\begin{equation}
 \delta^- \equiv p_{b=\ua}^- - p_{b=\da}^-.
\end{equation}
Here, $p_{b=\sigma}^-$ denotes the probability that an \emph{incoming} bit is in state $\sigma\in\{\ua,\da\}$. 
Analogously, $\delta^+ \equiv p_{b=\ua}^+ - p_{b=\da}^+$ quantifies the excess of spin up electrons 
in the \emph{outgoing} tape. 
As with the quantum dot spin valve, we assume that the spin up and spin down states on the bit dots 
are energetically degenerate, i.e., $H_B = \epsilon_b(|\ua\rangle\langle\ua| + |\da\rangle\langle\da|)$. 
A visualization of the system and bit string is provided in Fig. \ref{fig sketches sys ratchet} (a). 

\subsection{System-Bit Interaction}
\label{sec system bit interaction}

Finally, we must specify how the system and the bit interact during the short time $\delta t$. 
We will model the interaction between a filled quantum dot and the $n$'th bit by an effective Heisenberg Hamiltonian of the form 
$J_n(t) \boldsymbol\sigma_S\cdot\boldsymbol\sigma_{B_n}$.
Here $\boldsymbol\sigma = (\sigma^x,\sigma^y,\sigma^z)^T$ is the vector of Pauli matrices and 
$J_n(t) \ge 0$ is a coupling strength whose time dependence is induced by proximity between
the system and bit. If the interaction occurs around the time $t=t_n$, we demand that 
\begin{equation}\label{eq coupling J}
 \int_{-\infty}^\infty ds J_n(s) \approx \int\limits_{t_n-\delta t/2}^{t_n+\delta t/2} ds J_n(s) = \frac{\pi}{4},
\end{equation}
i.e., $J_n(t)$ is approximately zero for $t\notin[t_n-\delta t/2,t_n+\delta t/2]$. Our assumption 
$\delta t\ll\Gamma^{-1}$ allows us to neglect any dissipative dynamics arising from the interaction with the leads.
The time evolution of the combined system and bit -- given that the spin valve quantum dot is initially filled with an 
electron -- is governed by the unitary operator 
\begin{equation}
 \exp\left[-i\int\limits_{t_n-\delta t/2}^{t_n+\delta t/2} ds J_n(s) \boldsymbol\sigma_S\cdot\boldsymbol\sigma_{B_n}\right] \sim U_{\text{swap}}.
\end{equation}
Here, $U_\text{swap}$ is the unitary swap operator acting on two spins as 
$U_\text{swap}|\sigma_1\sigma_2\rangle = |\sigma_2\sigma_1\rangle$. Thus, if the spins of the system and the bit are the 
same then nothing happens, but if they are opposite then they both flip. If the quantum dot of the spin valve is initially 
empty, $|0\rangle$, we assume that the state of the bit does not change during the interval $t_n-\delta/2 < t < t_n+\delta/2$.

The exact form of the time dependent coupling strength $J_n(t)$ 
is unimportant: as long as $\delta t \ll \Gamma^{-1}$, it only needs to fulfill Eq. (\ref{eq coupling J}). 
In practice, of course, $J_n(t)$ will never exactly fulfill Eq. (\ref{eq coupling J}) but will differ by a small amount $\epsilon$ 
such that the true unitary operation we implement is given by $U_\text{swap} + \epsilon U_\text{cor} + \C O(\epsilon^2)$ 
where the correction term $U_\text{cor}$ gives rise to quantum coherences between the spin up and down states and 
a finite probability proportional to $\epsilon^2$ that the swap 
operation does not succeed. Because our treatment is in any case idealized (for instance, due to the assumed perfect 
polarization of the leads and the absence of phonons) we will also assume that $\epsilon = 0$, but a more 
sophisticated treatment would include a discussion about the so called \emph{fidelity} of our \emph{quantum gate} 
(i.e., our swap operation). 

The effective Heisenberg Hamiltonian above can be physically more rigorously justified by considering a 
Hubbard Hamiltonian for the two quantum dots (spin valve plus bit), introducing a weak electron tunneling term between the dots 
and for which one can show that the effective 
low energy interaction is governed by the Heisenberg Hamiltonian above \cite{FuldeBook}. The fact that this 
allows the realization of a swap gate is also used for purposes of quantum information processing 
in quantum dots \cite{LossDiVincenzoPRA1998}. In these setups the time dependent coupling $J_n(t)$ is induced by an 
electrostatic barrier, which can be controlled by a gate voltage, between two spatially fixed neighbouring quantum dots. 
Thus, we do not necessarily need a tape of bits moving past the spin valve but we could also use a single electron transistor 
or any other single electron source adjacent to the spin valve and control the interaction by a gate voltage. 
However, for similarity with the models proposed in Refs. 
\cite{MandalJarzynskiPNAS2012, BaratoSeifertEPL2013, HorowitzSagawaParrondoPRL2013, MandalQuanPRL2013, DeffnerJarzynskiPRX2013, DeffnerPRE2013, HoppenauEngelEPL2014} 
we will stick to the picture of a tape of quantum dots pulled along the spin valve. 
Finally, note that a SWAP operation between two spin qubits in quantum dots was experimentally realized already in 
2005 \cite{PettaEtAlScience2005}. 

\subsection{System Dynamics}

We are now ready to formulate the dynamical evolution of the system over one cycle of duration $\tau$.
For conciseness we henceforth neglect the small time window $\delta t$ and simply assume that the swap operation happens instantaneously.
We start a cycle with the system bit interaction, eventually leading to a swap operation, and end a cycle shortly before 
the next bit comes to interact with the system. Once we have reached a periodic steady state we can drop the 
index $n$ specifying the bit of the tape because the statistical behaviour of the system is the same from one interaction interval to the next. 
For the interval during which there is no system bit interaction, the time evolution of the system is generated by the Liouvillian (\ref{eq Liouvillian spin valve}), i.e., 
\begin{equation}\label{eq system time evolution 1}
 \rho_S^-(t+\tau) = e^{\C W\tau}\rho_S^+(t),
\end{equation}
where the superscript $-$ ($+$) denotes the state of the system before (after) the swap operation. 
The system state after the swap operation is related to the system state before via the transformation 
\begin{equation}\label{eq system time evolution 2}
 \rho_S^+ = \mbox{tr}_B[U\rho_S^-\otimes\rho_{B}^-U^\dagger] = 
 \left(\begin{array}{c}
        p_0^-	\\	p_1^-\frac{1+\delta^-}{2}	\\	p_1^-\frac{1-\delta^-}{2}	\\
       \end{array}\right)
\end{equation}
where $U \equiv 1_B\otimes|0\rangle_S\langle0| + U_\text{swap}|1\rangle_S\langle1|$ and 
$|1\rangle_S\langle1| = |\ua\rangle_S\langle\ua| + |\da\rangle_S\langle\da|$ denotes the projector onto the 1-electron subspace in the 
spin valve. We have introduced the notation $p_1^- \equiv p_\ua^- + p_\da^-$ and have used the fact that the initial state of the bit is given by 
$\rho_B^- = p_{b=\ua}^-|\ua\rangle_B\langle\ua| + p_{b=\da}^-|\da\rangle_B\langle\da|$ with $p_{b=\ua}^- = (1+\delta^-)/2$ and 
$p_{b=\da}^- = (1-\delta^-)/2$. 

Eq. (\ref{eq system time evolution 1}) together with Eq. (\ref{eq system time evolution 2}) 
fully specify the time evolution of the \emph{system}, which can now be solved for its periodic steady state. 
Because this expression is a bit lengthy we will not display it here. Furthermore, we will henceforth measure $\tau$ in units 
of $\Gamma^{-1}$, i.e., we can set without loss of generality $\Gamma\equiv 1$. 

\subsection{Mapping to a Brownian Ratchet}

Although our model is taken from the field of electron transport, we will now show that it can indeed be mapped to a Brownian ratchet. 
This makes it easier to relate our model to other work in the field of information thermodynamics and Maxwell's demon where many models are based 
on classical Brownian dynamics, see for instance Ref. \cite{HorowitzSagawaParrondoPRL2013}, which describes a very similar device and 
Ref. \cite{ToyabeEtAlNatPhys2010} for an experimental realization. 

To see why this can be done, let us imagine that we have a particle hopping between three discrete states labeled $\ua,0,\da$ with corresponding 
energies $E_\ua = -V, E_0 = -\frac{V}{2}, E_\ua = 0$. If this particle is immersed in a heat bath, its thermally activated transitions
$\ua\leftrightarrow0$ and $0\leftrightarrow\da$ (note that the direct transition $\ua\leftrightarrow\da$ is forbidden) obey 
local detailed balance exactly as in Eq. (\ref{eq local detailed balance}). 
This picture is also supported by the fact that the stationary state of the spin valve, Eq. (\ref{eq steady state spin valve}), 
equals a canonical equilibrium state with respect to the energies defined above. 

The analogy to a Brownian ratchet demon, in which a particle is transported against an external force (or \emph{load}) 
by switching between two different potentials, is completed by recognizing that our fictitious particle could live on a discrete lattice 
$(\dots,\ua,0,\da,\ua,0,\da,\dots)$ with impenetrable walls between the states $\da$ and $\ua$, but is allowed to change 
between two different modes due to the interaction with a bit (or later due to the feedback, Sec. 
\ref{sec measurement based feedback scheme}). This is summarized in Fig. \ref{fig sketches sys ratchet} (b). 
It is easy to show that the work $\Delta W$ against the external load over one cycle is given by 
$\Delta W = -\frac{V}{2}\Delta N^L + \frac{V}{2}\Delta N^R = -V\Delta N^L$ where $\Delta N^\nu$ denotes the average number of 
tunneled particles from lead $\nu$ during one cycle and we have used particle number conservation $\Delta N^R = -\Delta N^L$. 
We will define $\Delta N^\nu$ more thoroughly in Sec. \ref{sec work heat bit statistics}.

\section{Thermodynamics}
\label{sec thermodynamics}

\begin{figure}
\includegraphics[width=0.48\textwidth,clip=true]{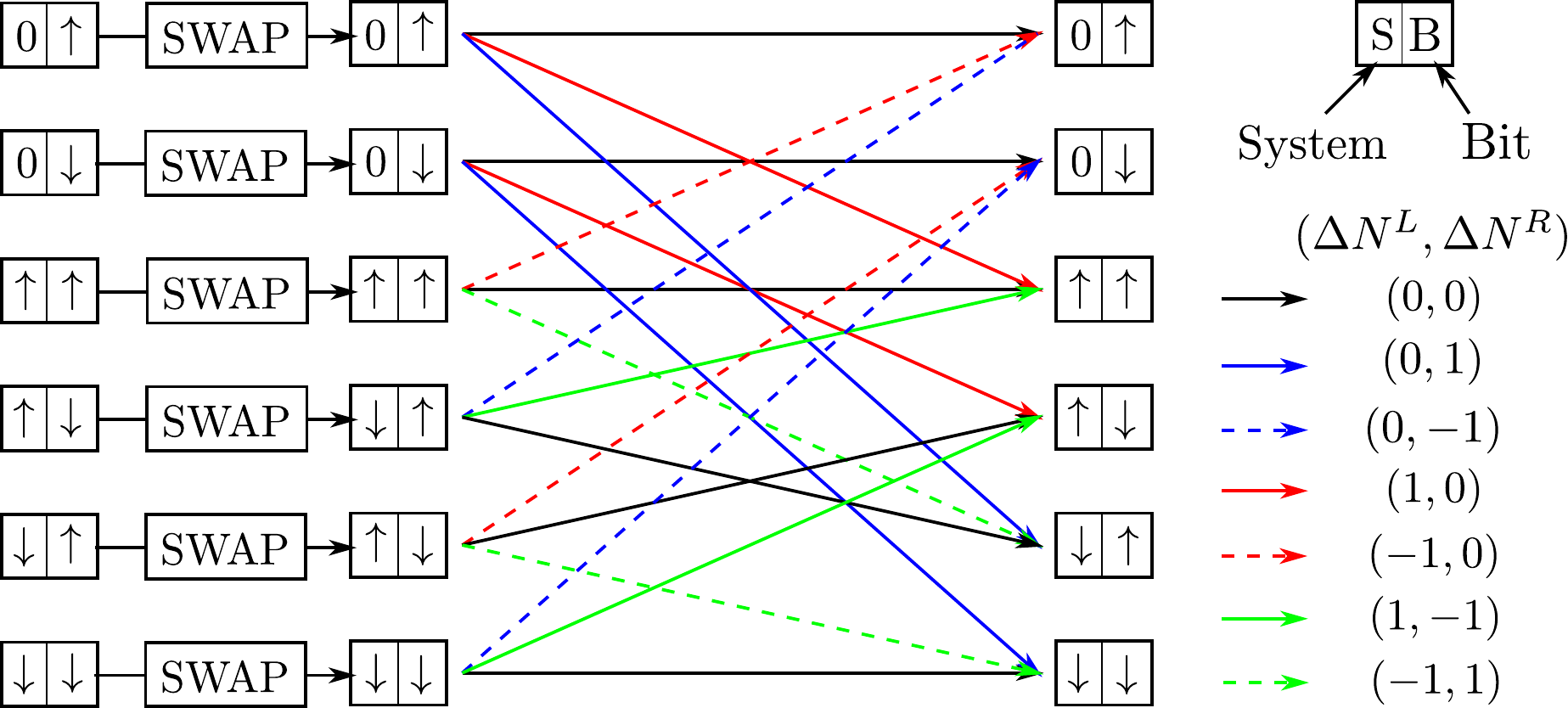}
\caption{\label{fig transitions} Sketch of the possible transitions during a single cycle, starting from the left with 
the initial state $\rho_S^-\otimes\rho_B^-$, followed by a swap operation, and ending after a time $\tau$ in the final state, 
which becomes the new initial state for the next cycle if we replace it with a new bit. The arrows indicate possible transitions 
and the color code symbolizes the net number of exchanged particles with the left and right reservoir contributing to Eqs. 
(\ref{eq particle number L}) and (\ref{eq particle number R}). }
\end{figure}

Before we start with the thermodynamic analysis we make two important remarks: first, the full behaviour of the model 
depends only on three parameter: the voltage $V\in(-\infty,\infty)$, which favors a particular spin direction in the spin valve, the excess parameter 
$\delta^-\in[-1,1]$ of the incoming bits and the interval duration $\tau\in(0,\infty)$. Second, our system possesses a symmetry:
mapping the parameters $(V,\delta^-)$ to $(-V,-\delta^-)$ is physically equivalent to exchanging the labels $L$ and $R$ and 
$\ua$ and $\da$ of the spin valve. When we later plot the phase diagram of our model over 
$V\times\delta^- = (-\infty,\infty)\times[-1,1]$ this symmetry allows us to limit the discussion 
to one half of the phase space because the behaviour in the other half follows by symmetry arguments. 

\subsection{Work, Heat and Bit Statistics}
\label{sec work heat bit statistics}

In appendix \ref{sec app Proof Delta W zero} we show that the work expenditure $\Delta W_\text{pull}$ to pull the tape 
in our idealized setup (assuming a frictionless sliding tape) is truly zero.  Physically, this is due to the two facts that we treat 
the swap operation unitarily and that the spin up and down states are energetically degenerate 
in the system and bit quantum dot. 

Next, we determine the flow of electrons at the left or right contact, defined to be positive 
if the electron hops out of the reservoir into the system.
This is most easily done by looking at all possible transitions during one cycle and associating 
the corresponding change in particle number with it as shown in Fig. \ref{fig transitions}. For instance, for 
the number of tunneled particles at the left contact we must add the terms (following the diagram in Fig. 
\ref{fig transitions} from top to bottom) 
\begin{equation}
 \begin{split}
  \Delta N^L	=&~	p_0^- p_{b=\ua}^- P_{0\ra\ua} + p_0^-p_{b=\da}^- P_{0\ra\ua}	\\
		&-	p_\ua^-p_{b=\ua}^- (P_{\ua\ra0} + P_{\ua\ra\da}) + p_\ua^-p_{b=\da}^- P_{\da\ra\ua}	\\
		&-	p_\da^-p_{b=\ua}^- (P_{\ua\ra0} + P_{\ua\ra\da}) + p_\da^-p_{b=\da} P_{\da\ra\ua}.
 \end{split}
\end{equation}
Here, $P_{s\rightarrow s'}$ denotes the transition probability from the system state $s$ to $s'$ due to 
the Liouvillian $\C W$. For $\tau\rightarrow\infty$ this becomes $P_{s\rightarrow s'} = \bar p_{s'}$ 
with $\bar p_{s'}$ from Eq. (\ref{eq steady state spin valve}). Parametrizing the bit probabilities by the 
excess parameter $\delta^-$ yields after some manipulation for $\Delta N^L$ and, following a similar 
procedure, for $\Delta N^R$: 
\begin{align}
 \Delta N^L	=&~	p^-_{0}P_{0\rightarrow\ua}	\label{eq particle number L}	\\
		&+	p^-_{1}\left[\frac{1-\delta^-}{2} P_{\da\rightarrow\ua} - \frac{1+\delta^-}{2} \big(P_{\ua\rightarrow0} + P_{\ua\rightarrow\da}\big)\right],	\nonumber	\\
 \Delta N^R	=&~	p^-_{0}P_{0\rightarrow\da}	\label{eq particle number R}	\\
		&+	p_{1}^-\left[\frac{1+\delta^-}{2} P_{\ua\rightarrow\da} - \frac{1-\delta^-}{2} \big(P_{\da\rightarrow0} + P_{\da\rightarrow\ua}\big)\right].	\nonumber
\end{align}
It is possible to verify particle 
number conservation $\Delta N^L + \Delta N^R = 0$ and for later purposes we note the following identity: 
\begin{equation}\label{eq help useful identity}
 2\Delta N^L = \Delta N^L-\Delta N^R = (p_{0}^- - 1)\delta^- + p_{\ua}^- - p_{\da}^-.
\end{equation}
The average current is defined by $I^L(\tau) \equiv \frac{\Delta N^L(\tau)}{\tau}$. 
Two particularly simple expressions are obtained in the limits 
\begin{equation}
 \lim_{V\rightarrow0}\lim_{\tau\rightarrow\infty} \Delta N^L = -\frac{\delta^-}{3}, ~~~ \lim_{V\rightarrow0}\lim_{\tau\rightarrow0} I^L = -\frac{\delta^-}{6}.
\end{equation}
The minus sign indicates that an excess of spin up electrons on the tape ($\delta^->0$) 
generates a current from right to left. 

Finally, we compute the change in the statistics of the outgoing bits. The excess parameter 
of the tape after the interaction with the system is given by 
\begin{equation}\label{eq delta plus tape}
 \delta^+ = p_{b=\ua}^+ - p_{b=\da}^+ = p_{0}^- \delta^- + p_{\ua}^- - p_{\da}^-.
\end{equation}
Using Eq. (\ref{eq help useful identity}) we can write 
\begin{equation}\label{eq help useful identity 2}
 \delta^+ = \delta^- + 2\Delta N^L,
\end{equation}
an identity we will use later on prove a second law like inequality \footnote{Also compare this expression 
with the equivalent Eqs. [S17] and [3] in Ref. \cite{MandalJarzynskiPNAS2012}. The equivalence is obvious 
by noting that the circulation $\Phi$ in Ref. \cite{MandalJarzynskiPNAS2012} corresponds to the change 
in particle number $\Delta N^L$ in our case, where $\Phi$ is proportional to the extracted work $W$ 
in Ref. \cite{MandalJarzynskiPNAS2012} in the same way as $\Delta N^L$ is proportional to the extracted work 
in our Brownian ratchet version. }. Having $\delta^+$ allows us to compute the change in the Shannon entropy 
of the bits defined by \footnote{Note that this definition ignores correlations in the outgoing bit stream, which 
are present but rather small in our case. For $\tau\rightarrow\infty$ these correlations vanish, because the 
system has time to relax to the steady state (\ref{eq steady state spin valve}), which is uncorrelated from the 
state of the bit. Also see, for instance, the discussion in Ref. \cite{MandalJarzynskiPNAS2012}. }
\begin{equation}\label{eq change Shannon entropy bits}
 \Delta H_B \equiv H[\delta^+] - H[\delta^-]
\end{equation}
with 
\begin{equation}\label{eq Shannon entropy bits}
 H[\delta] \equiv -\frac{1+\delta}{2}\ln\frac{1+\delta}{2} - \frac{1-\delta}{2}\ln\frac{1-\delta}{2}.
\end{equation}
If $\Delta H_B>0$ we effectively \emph{write} information to the tape, i.e., increase its entropy, 
while if $\Delta H_B<0$ implies that we \emph{erase} information from the tape effectively lowering its entropy. 

\subsection{Working Modes of the System}

\begin{figure}
\includegraphics[width=0.42\textwidth,clip=true]{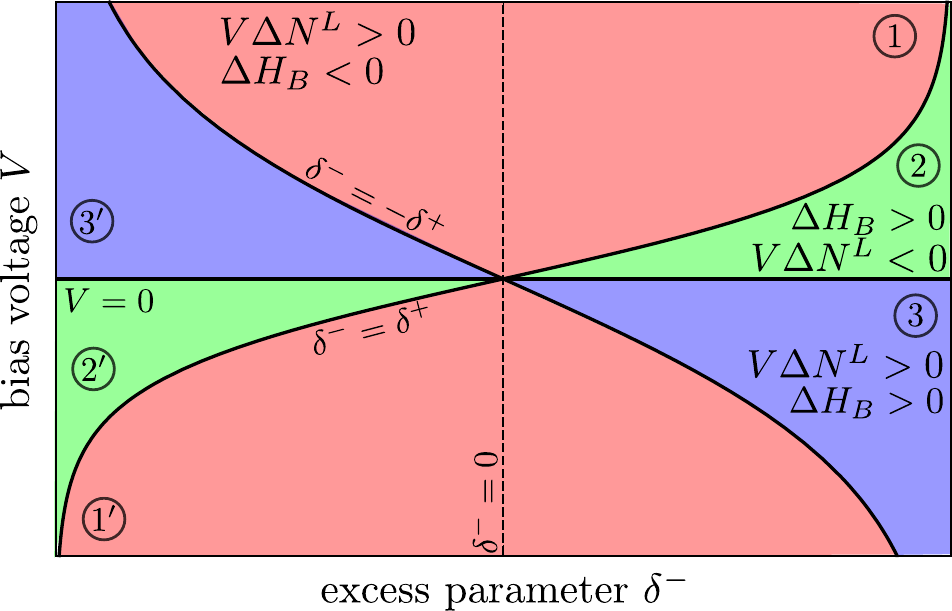}
\caption{\label{fig phase diagram} Phase diagram of the system over $\delta^-$ and $V$ showing the different 
working modes: information eraser mode (regions 1 and $1^\prime$, red color), Maxwell demon mode (regions 2 and $2^\prime$, 
green color) and the `third mode' (regions 3 and $3^\prime$, blue color). The primed regions are connected to the 
unprimed regions by the symmetry relation. For clarity we also plot the line $\delta^- = 0$ 
(dashed). Note that the line $\delta^+=-\delta^-$ changes with $\tau$ but the overall structure remains the 
same.}
\end{figure}

In principle, one can distinguish three different working modes in which our device can operate. 
First, it can be used to pump electrons against the bias (i.e., $V\Delta N^L<0$) while simultaneously 
writing information on the tape ($\Delta H_B>0$) what we will call the `Maxwell demon mode'. 
Second, one can use it to erase information ($\Delta H_M<0$) by exploiting the flow of electrons 
with the bias ($V\Delta N^L>0$) (`eraser mode'). Third, we can also have $V\Delta N^L>0$ and $\Delta H_B>0$ 
where electrons flow with the bias while simultaneously information is written 
on the tape \footnote{Note that we always have $V\Delta N^L = 0$ if the spin valve does not interact with a tape. 
Thus, in contrast to Ref. \cite{MandalJarzynskiPNAS2012} it is probably better not to call this third mode 
a `dud' since it can still accomplish a useful task.}. 
A possible working mode where $V \Delta N^L<0$ and $\Delta H_B<0$ is forbidden by the second law 
of thermodynamics as we will see in the next section. 

These modes are summarized in Fig. \ref{fig phase diagram} and are separated from each other by three lines in the phase diagram.
One line is characterized by $V=0$ and the other two lines are given by $\Delta H_B = 0$, which can happen 
either if $\delta^+ = \delta^-$ or if $\delta^+ = -\delta^-$. The first case is given by 
\begin{equation}
 \delta^+ = \delta^- \Leftrightarrow \delta^- = \frac{e^V-1}{e^V+1} = \tanh\frac{V}{2}
\end{equation}
where $\tanh(x)$ denotes the hyperbolic tangent. This line is independent of $\tau$. By contrast, 
the line for which $\delta^+ = -\delta^-$ depends on $\tau$ and has in general a more complicated 
functional relation. 

The total phase space $V\times \delta^-$ for a given $\tau$ is divided into six regions, of which only 
three are independent due to the symmetry. Specifically, we have the relations 
\begin{align}
 \Delta N^L(V,\delta^-)	&=	\Delta N^R (-V,-\delta^-),	\\
 \delta^+(V,\delta^-)	&=	-\delta^+(-V,-\delta^-),	\\
 \Delta H_B(V,\delta^-)	&=	\Delta H_B(-V,-\delta^-).
\end{align}

\subsection{Second Law}

In appendix \ref{sec app Proof 2nd law tape model} we obtain the following second law like inequality: 
\begin{equation}\label{eq 2nd law tape model}
 \Delta_{\bb i} S_\text{tape} \equiv V\Delta N^L + \Delta H_B = -\Delta W + \Delta H_B \ge 0
\end{equation}
using the equivalence between the spin valve and the Brownian ratchet model ($V\Delta N^L = -\Delta W$). 
We will call $\Delta_{\bb i} S_\text{tape}$ the entropy production (per cycle) of the tape model. 
In fact, from the point of view of the system there are different versions of the second law possible 
as we will see in the next section. Thus, it is hard to justify speaking about \emph{the} second law for our device. 

\section{Measurement based Feedback Scheme}
\label{sec measurement based feedback scheme}

We now turn to a closed-loop or feedback control scheme, which relies on explicit measurements 
and subsequent control actions depending on the measurement outcomes. This approach is closer to 
the orginal idea of Maxwell. In fact, our inclusive approach above has nothing to do with \emph{feedback} 
control because we do not use the information we are writing on the tape to change the system dynamics. 
Moreover, the `information' on the tape does not directly represent information about the state of the system, 
but only about $\Delta N_L$, a quantity which is rather related to the reservoir than to the system. 

\subsection{Measurement and Feedback}
\label{sec measurement and feedback}

\begin{figure}
\includegraphics[width=0.42\textwidth,clip=true]{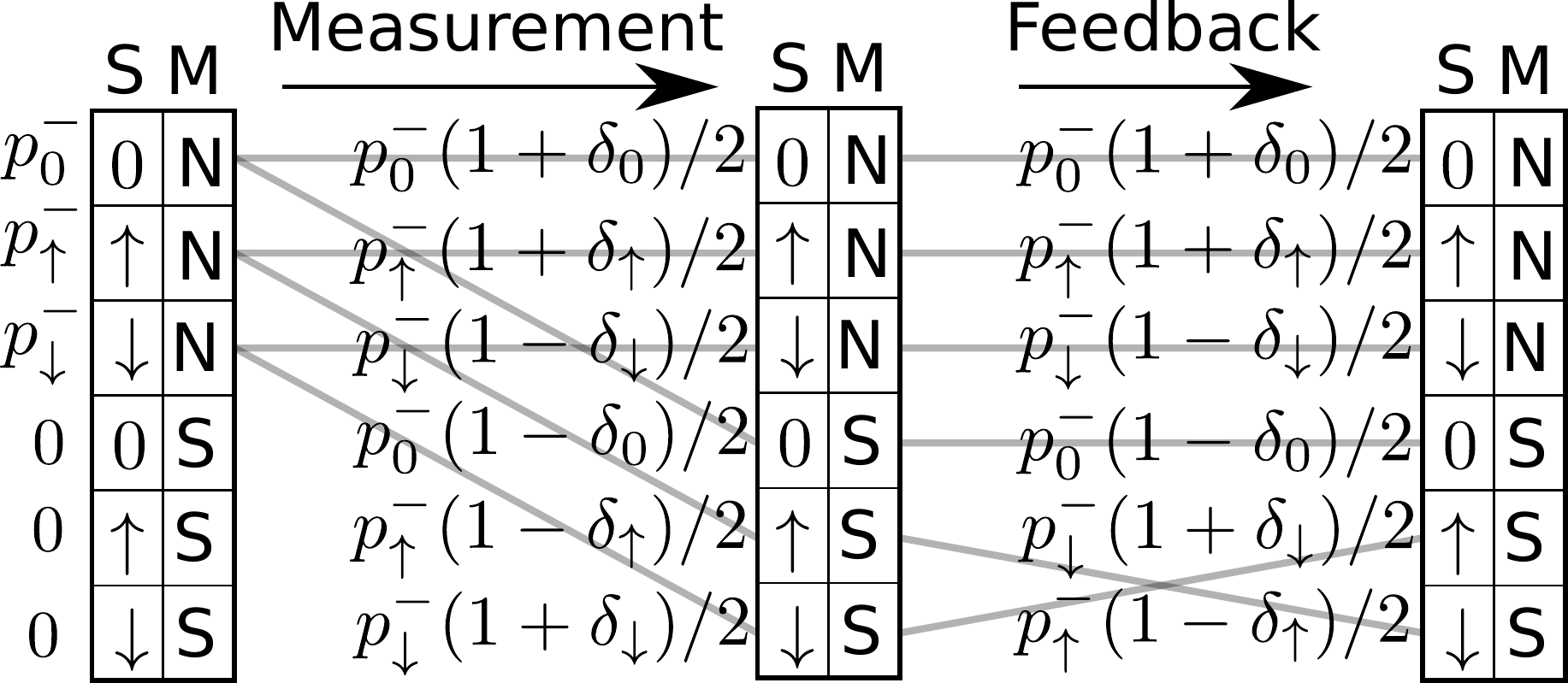}
\caption{\label{fig meas and fb} Probabilities to find the system and the memory in their respective states 
prior to the measurement, after the measurement and after the feedback. The possible transitions are marked 
with shaded grey lines. }
\end{figure}

We assume we have a 1-bit memory $M$ with state space $\C M = \{N,S\}$ where $N,S$ will specify the action 
of the feedback later on \footnote{It is easy to think of memories with more than two states, but in 
a certain sense a two state memory is minimal and could be also realized by a bit in our tape model. 
Thus, we will focus on the two state memory.}. 
Prior to the measurement the combined probability distribution is given by 
$p(s,m) = p(s) p(m) = p_s^- \delta_{mN}$ where we assume that the state of the memory is initially in $m=N$ 
and we denote $p(s) = p_s^-$ as in the previous section. We then perform a measurement of the quantum dot 
spin valve: ideally, if the state of the dot is $s\in\{0,\ua\}$ we leave the memory in the state $N$, but if 
the state of the dot is $s=\da$ we set the memory to the state $S$. In general there will be a measurement error, 
which can be characterized by the conditional probability $P[m|s]$ to find the memory in the state $m$ given the 
system was in state $s$. The state after the measurement is then given by $p'(s,m) = P[m|s] p_s^-$ with 
$\sum_m P[m|s] = 1$. We parametrize the conditional probabilities as \footnote{Very often one uses a 
parameter $\epsilon_s\in[0,1]$ ($s\in\{0,\ua,\da\}$) to quantify the measurement error, but for better comparison 
with the discussion in Sec. \ref{sec thermodynamics} we use the excess parameter $\delta_s \equiv 1-2\epsilon_s$ 
instead, which now quantifies the excess of correct measurements compared to faulty ones (for $\delta_s>0$).}
\begin{equation}\label{eq cond prob measurement}
 P[S|0] = \frac{1-\delta_0}{2}, ~ P[S|\ua] = \frac{1-\delta_\ua}{2}, ~ P[N|\da] = \frac{1-\delta_\da}{2}. 
\end{equation}

Finally, after the measurement we flip the spin of the quantum dot, if the memory is in state $S$ (`switch') or 
we do nothing if the memory is in state $N$ (`no switch'). Ideally, because the energy for spin up and down in 
the quantum dot are assumed to be equal, this feedback operation requires no 
energy expenditure. Of course, if the state of the system is $0$, then nothing happens. The combined probability 
distribution of system and memory after the feedback can be inferred from Fig. \ref{fig meas and fb}. 
The state of the quantum dot spin valve is now given by 
\begin{equation}\label{eq system state after feedback}
 \rho^+_S 
 = \left(\begin{array}{c}
          p_{0}^-	\\	p_{\ua}^-\frac{1+\delta_\ua}{2} + p_{\da}^-\frac{1+\delta_\da}{2}	\\	p_{\da}^-\frac{1-\delta_\da}{2} + p_{\ua}^-\frac{1-\delta_\ua}{2}	\\
         \end{array}\right).
\end{equation}
To obtain identical system dynamics compared to the situation involving the tape of bits, we only have to assure 
that Eq. (\ref{eq system time evolution 2}) coincides with Eq. (\ref{eq system state after feedback}) 
because the time evolution in between two  successive measurement and feedback steps is the same as before, see 
Eq. (\ref{eq system time evolution 1}). Thus, we will from now on choose 
\begin{equation}\label{eq meas error}
 \delta_\ua = \delta_\da \equiv \delta^-.
\end{equation}
All quantities associated to the system (such as $\Delta N^L$) are the same for both approaches. What makes the two 
approaches different is the form of information processing and the related second laws. 

\subsection{Second Law}

The second law of thermodynamics for feedback controlled systems predicts 
\cite{CaoFeitoPRE2009, DeffnerLutzArXiv2012, HorowitzSagawaParrondoPRL2013} 
\begin{equation}\label{eq 2nd law feedback}
 \Delta_{\bb i} S_\text{fb} \equiv V\Delta N^L + I(S;M') \ge 0,
\end{equation}
which is equivalent to $\Delta W \le I(S;M')$ due to our mapping to the Brownian ratchet model. 
Here, $I(S;M')$ is the mutual information, which measures the amount of correlation between the system state 
before the feedback and the post-measurement state of the memory. It can be defined as \cite{CoverThomasBook} 
\begin{equation}
 I(S;M') \equiv H_{M'} - H_{M'|S}
\end{equation}
where $H_{Y|X} \equiv -\sum_x p(x) \sum_y p(y|x)\ln p(y|x)$ is the conditional entropy of $Y$ given $X$. 
The mutual information is always non-negative, $I(X;Y)\ge0$, and vanishes only 
if $p(x,y) = p(x)p(y)$ \cite{CoverThomasBook}. 

More explicitly, the mutual information can be written in our case as 
\begin{equation}
 \begin{split}
  I(S;M')	=&	-\sum_m p'(m)\ln p'(m)	\\
		&+	\sum_s p_s^- \sum_m P[m|s]\ln P[m|s]
 \end{split}
\end{equation}
where $p'(m) = \sum_s p'(m,s) = \sum_s P[m|s]p_s^-$ is the marginal probability distribution of the memory 
after the measurement and the conditional probabilities $P[m|s]$ were introduced in Eq. 
(\ref{eq cond prob measurement}), see also Fig. \ref{fig meas and fb}. Together with Eq. 
(\ref{eq meas error}) and parametrizing the marginal probability distribution as 
$p'(m=N) = (1+\delta_M)/2$ and $p'(m=S) = (1-\delta_M)/2$, we can write the mutual 
information as 
\begin{equation}\label{eq mutual information}
 I(S;M') = H[\delta_M] - p_0^- H[\delta_0] - p_1^- H[\delta^-]
\end{equation}
with $H[x]$ defined as in Eq. (\ref{eq Shannon entropy bits}). 
Here, $\delta_M$ is given by 
\begin{equation}\label{eq delta M feedback}
 \delta_M = \delta_0 p_0^- + \delta^-(p_\ua^- - p_\da^-).
\end{equation}
This equation is analogous to Eq. (\ref{eq delta plus tape}).

\subsection{Comparison of the two Second Laws}
\label{sec comparison two second laws}

\begin{figure}
\includegraphics[width=0.45\textwidth,clip=true]{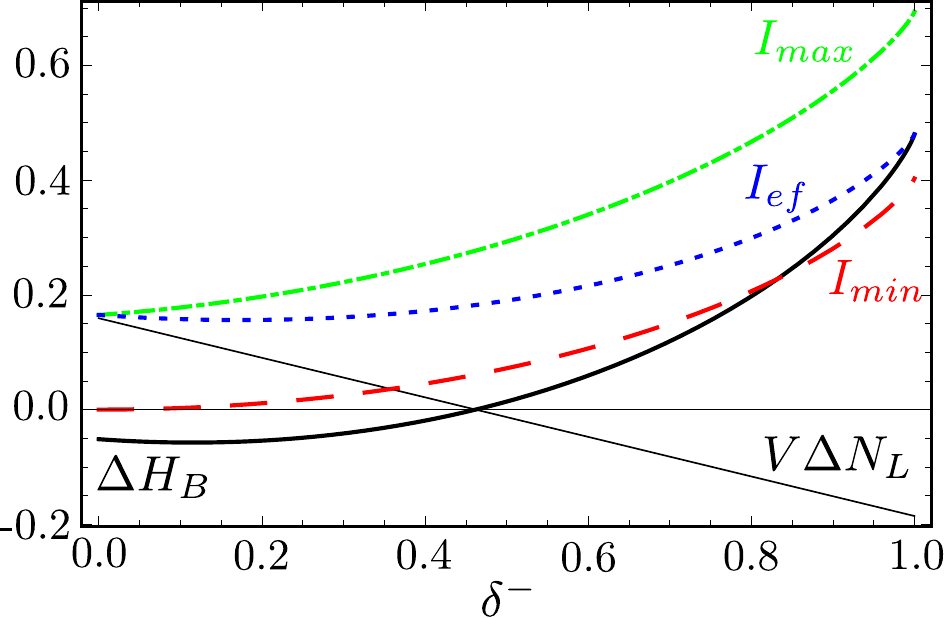}
\caption{\label{fig inclusive vs feedback} Plot of the delivered work $-\Delta W = V\Delta N^L$ (black thin line), 
the change in the Shannon entropy of the bits $\Delta H_B$ (black thick line) and the maximum $I_\text{max}$ 
(green dash-dotted line), the minimum $I_\text{min}$ (red dashed line) and the error-free mutual information 
$I_\text{ef}$ (blue dotted line) over the excess parameter $\delta^-$. We see that the sum of $V\Delta N^L$ and any 
of the information theoretic quantities $\Delta H_B, I_\text{max}, I_\text{min}$ or $I_\text{ef}$ is always positive 
as imposed by Eqs. (\ref{eq 2nd law tape model}) and (\ref{eq 2nd law feedback}). We further chose $V=1$ and 
$\tau = 10$. }
\end{figure}

We wish to compare the second law from our inclusive approach, Eq. (\ref{eq 2nd law tape model}), with the second 
law for feedback controlled system, Eq. (\ref{eq 2nd law feedback}), using the mutual information from 
Eq. (\ref{eq mutual information}). First of all, we note that one can not address problems like information erasure 
within the feedback approach because the mutual information is always positive. Hence, we will focus 
on the Maxwell demon regime, i.e., region 2 in Fig. \ref{fig phase diagram}. 
Furthermore, we see that the mutual information still depends on 
$\delta_0$ although the dynamics of the system are completely \emph{independent} of $\delta_0$. 
For a better comparison we thus find it sensible to compare the case $\delta_0=1$, which corresponds to an 
error-free measurement of the empty dot state, with the maximum and minimum value of the mutual 
information in dependence of the parameter $\delta_0$. These are attained for 
\begin{equation}\label{eq max min delta0}
 \delta_0^\text{max} = -1, ~~~ \delta_0^\text{min} = \delta^-\frac{p_\ua^- - p_\da^-}{p_\ua^- + p_\da^-}
\end{equation}
as we establish in appendix \ref{sec app max min mutual info}. The mutual information (\ref{eq mutual information}) 
in these three cases becomes after using Eq. (\ref{eq delta M feedback}) 
\begin{align}
 I_\text{max}	&\equiv	I(S;M')(\delta_0^\text{max})	\\
		&=	H[-p_0^- + \delta^-(p_\ua-p_\da)] - p_1^-H[\delta^-],	\nonumber	\\
 I_\text{ef}	&\equiv	I(S;M')(\delta_0=1)	\\
		&=	H[p_0^- + \delta^-(p_\ua-p_\da)] - p_1^-H[\delta^-],	\nonumber	\\	
 I_\text{min}	&\equiv	I(S;M')(\delta_0^\text{min})	\\
		&=	H[\delta_0^\text{min}] - p_0^-H[\delta_0^\text{min}] -  p_1^-H[\delta^-]	\nonumber
\end{align}
where the subscript `ef' stands for `error-free'. 
The maximum amount of correlation between system and memory is thus reached for $\delta_0=-1$, which indeed would also 
correspond to an error-free measurement because a measurement, which always yields the wrong outcome, is very reliable, 
too. On the other hand, the dynamically irrelevant parameter $\delta_0$ 
gives us the possibility to decrease the amount of correlation such that we obtain a stronger bound in Eq. 
(\ref{eq 2nd law feedback}) on the maximum extractable work $\Delta W$. 

Numerical results are presented in Fig. \ref{fig inclusive vs feedback} showing that neither the second law for the 
inclusive approach nor the second law for feedback control always provides a tighter bound on the maximum amount of 
extractable work. Thus, there is probably no trivial relationship connecting the two different second laws for 
arbitrary $\delta^-$. Even for the special case of a fully ordered tape, i.e., for $\delta^-=1$, there seems to be no 
general relation, but we can easily confirm $I_\text{min} \le \Delta H_B = I_\text{ef} \le I_\text{max}$. 

\section{Effective Master Equation}
\label{sec effective master equation}

\subsection{Poisson distributed Tape}
\label{sec Poisson distributed tape}

Previously the interaction with the bit occurred at regular intervals $\tau$, i.e., the so called waiting 
time distribution was given by $p_t = \delta(t-\tau)$. Instead, we now assume an exponential waiting time 
distribution $p_t = \gamma e^{-\gamma t}$, which implies that the number of events (interactions with 
a bit) in a fixed interval is Poisson distributed. Here, the parameter $\gamma>0$ describes the 
strength of the Poisson process. Although we speak of `interaction with a bit', the 
resulting ME for the system is the \emph{same} as if we would use the feedback scheme from the previous section 
with Poisson distributed measurement times. 

Due to the Poisson distributed tape we can now describe the evolution of the system by a single differential 
equation as opposed to the discrete maps (\ref{eq system time evolution 1}) and 
(\ref{eq system time evolution 2}). The resulting effective ME is derived in appendix 
\ref{sec app derivation eff ME} and can be written as $\partial_t\rho_S(t) = \C W_\text{eff}\rho_S(t)$ with 
$\C W_\text{eff} = \C W_L + \C W_R + \C W_B$ where $\C W_L$ and $\C W_R$ are given in Eq. 
(\ref{eq Liouvillian spin valve}) and 
\begin{equation}\label{eq effective ME}
 \C W_B = \frac{\gamma}{2}\left(\begin{array}{ccc}
				 0	&	0		&	0		\\
				 0	&	-(1-\delta^-)	&	1+\delta^-	\\
				 0	&	1-\delta^-	&	-(1+\delta^-)	\\
				\end{array}\right).
\end{equation}
The contribution $\C W_B$ due to the interaction with the information reservoir, i.e., the tape of bits, has an apparent 
non-thermal character 
and is responsible for modifications of thermodynamic relations even in the absence of information processing or 
feedback \cite{KanazawaSagawaPRE2013}. The effective rate matrix ${\cal W}_B$ could, however, also result from 
coarse-graining an underlying microscopic model, which has a standard thermodynamic interpretation (as in Ref. 
\cite{StrasbergSchallerPRL2013}). 

We see that the effective Liouvillian possesses a clear separation into contributions from three different 
`reservoirs', and thus can be examined using standard thermodynamics. In this picture, each reservoir 
induces certain transitions from $\sigma$ to $\sigma'$ with corresponding currents $I[\sigma\rightarrow\sigma']$. 
We abbreviate $I^L = I[0\rightarrow\ua], I^R = I[0\rightarrow\da]$ and 
$I^B = I[\ua\rightarrow\da] = \frac{\gamma}{2}[(1-\delta^-) p_\ua - (1+\delta^-) p_\da]$, 
which are related at steady state via 
\begin{equation}
 I^L + I^R = 0, ~~~ I^L - I^B = 0, ~~~ I^R + I^B = 0.
\end{equation}
Thus, it suffices to know only one current, e.g. $I^L$, to deduce the value of all other currents (`tight coupling'). 
Note that the heat flow between the bits and system is zero because the states $\ua$ and $\da$ have the same energy. 
Furthermore, $I^B$ describes the net rate of spin flips (from $\ua$ to $\da$) in the system, which equals the net 
rate of bit flips (now from $\da$ to $\ua$) in the tape. 

The entropy production associated with the rate matrix $\C W$ taking care of the splitting into the different reservoir 
contributions becomes 
\begin{equation}\label{eq 2nd law effective first}
 \dot S_{\bb i} = (V + \C A)I^L \ge 0, ~~~ \C A \equiv \ln\frac{1-\delta^-}{1+\delta^-}.
\end{equation}
Following Refs. \cite{EspositoSchallerEPL2012, StrasbergSchallerPRL2013, StrasbergSchallerPRE2013} we now 
interpret $\C A I^L$ as an effective information current because it shows up as a new term in the entropy 
production due to the interaction with the tape, and $\C A$ is an information or feedback affinity depending 
only on the parameters describing the state of the incoming bits (or the measurement error in the feedback scheme). 
A priori however it is not quite clear how the effective information current is related to information theoretic 
quantities as, e.g., entropy or mutual information.  While progress has been made toward clarifying this issue for 
bipartite systems \cite{HorowitzEspositoPRX2014, ShiraishiSagawaArXiv2014, DianaEspositoJSM2014, HartichBaratoSeifertJSM2014}, 
our model does not have a bipartite structure due to the simultaneous change of the system and bit state. Nevertheless, 
within our model, we can indeed show that (see appendix \ref{sec app proof of Eq}) 
\begin{equation}\label{eq equivalence effective and tape}
 \frac{dH_B(t)}{dt} = \C A I^L,
\end{equation}
that is to say the effective information current introduced \emph{ad hoc} coincides with the change of Shannon 
entropy of an idealized tape of bits, which induces the transitions between $\ua$ and $\da$. Thus, 
Eq. (\ref{eq 2nd law effective first}) becomes 
\begin{equation}\label{eq 2nd law effective}
 \dot S_{\bb i} = VI^L + \frac{dH_B(t)}{dt} \ge 0
\end{equation}
In this sense, the two second laws (\ref{eq 2nd law tape model}) and (\ref{eq 2nd law effective}) present the same 
physical statement. 

Note that the situation in which the effective ME (\ref{eq effective ME}) is valid is not as easy to compare with the 
previous situation of constant time intervals, as was the case for the two level system studied in 
Ref. \cite{BaratoSeifertPRL2014}. In Ref. \cite{BaratoSeifertPRL2014} the state of the system after the arrival 
of the new bit or after the feedback was \emph{independent} of the system state before. This allowed them to average 
quantities obtained for the constant interval case by the exponential distribution $\gamma e^{-\gamma t}$ to obtain 
the corresponding quantities for the Poisson distributed case. 
In contrast, in our case the 
state of the system after the swap operation or after the feedback is still dependent on the state of the system 
before, see Eqs. (\ref{eq system time evolution 2}) and (\ref{eq system state after feedback}). Thus, averaging 
previous quantities obtained in Secs. \ref{sec thermodynamics} and \ref{sec measurement based feedback scheme} 
with $\gamma e^{-\gamma t}$ is not meaningful, as they do not coincide with any of the quantities obtained from 
Eq. (\ref{eq effective ME}). 

However, we expect the case of constant intervals (Secs. \ref{sec thermodynamics} and 
\ref{sec measurement based feedback scheme}) to be comparable with the case of Poisson distributed intervals, 
if we let the size of the intervals go to zero, i.e., $\tau\rightarrow0$ or $\gamma\rightarrow\infty$. 
We will call this limit the `infinite fast feedback limit'. 

\subsection{Infinite fast Feedback Limit}
\label{sec infinite fast feedback limit}

In the infinite fast feedback limit it is again possible to derive a new effective Liouvillian 
$\tilde{\C W}_\text{eff} = \tilde{\C W}_L + \tilde{\C W}_R$ by coarse graining the fast dynamics 
associated with $\C W_B$. This new Liouvillian governs the time evolution of the reduced 
probability vector $(p_0,p_1)$ with $p_1 = p_\ua + p_\da$ and is given by 
\begin{equation}\label{eq effective effective ME}
 \begin{split}
  \tilde{\C W}_L	&=	\Gamma\left(\begin{array}{cc}
                     -f_L	&	(1-f_L)(1+\delta^-)	\\
                     f_L	&	-(1-f_L)(1+\delta^-)	\\
                    \end{array}\right),	\\
  \tilde{\C W}_R	&=	\Gamma\left(\begin{array}{cc}
                     -f_R	&	(1-f_R)(1-\delta^-)	\\
                     f_R	&	-(1-f_R)(1-\delta^-)	\\
                    \end{array}\right).
 \end{split}
\end{equation}
We have thus effectively traced out the interaction with the `bit reservoir' but obtained modified rates for 
the electrons to tunnel out of the dot. This modification of the rates, which yields a modified local detailed 
balance relation, was interpreted as a signature of an idealized Maxwell demon in 
\cite{SchallerEtAlPRB2011, EspositoSchallerEPL2012, StrasbergSchallerPRL2013}. 
We remark that it is also possible to deduce this ME directly, without invoking a Poisson distribution of the bits 
but by starting with Eqs. (\ref{eq system time evolution 1}) and (\ref{eq system time evolution 2}) and 
expanding the generator to first order in $\tau$: $e^{\C W\tau} = 1 + \C W\tau + \C O(\tau^2)$. 

The thermodynamic analysis of Eq. (\ref{eq effective effective ME}) again follows the standard procedure. We now 
have only two currents, which must balance: $\tilde I^L + \tilde I^R = 0$. Explicitly we have 
\begin{equation}
 \tilde I^L = -\frac{\delta^- + (\delta^- - 1)e^V + 1}{\delta^- - (\delta^- - 3)e^V + 6e^{V/2} + 3}.
\end{equation}
The second law takes on the 
same form as Eq. (\ref{eq 2nd law effective}), but is numerically different because it contains the current 
$\tilde I^L$ and not $I^L$: 
\begin{equation}\label{eq 2nd law effective effective}
 \dot{\tilde S}_{\bb i} = (V + \C A)\tilde I^L \ge 0.
\end{equation}

\begin{figure}
\includegraphics[width=0.45\textwidth,clip=true]{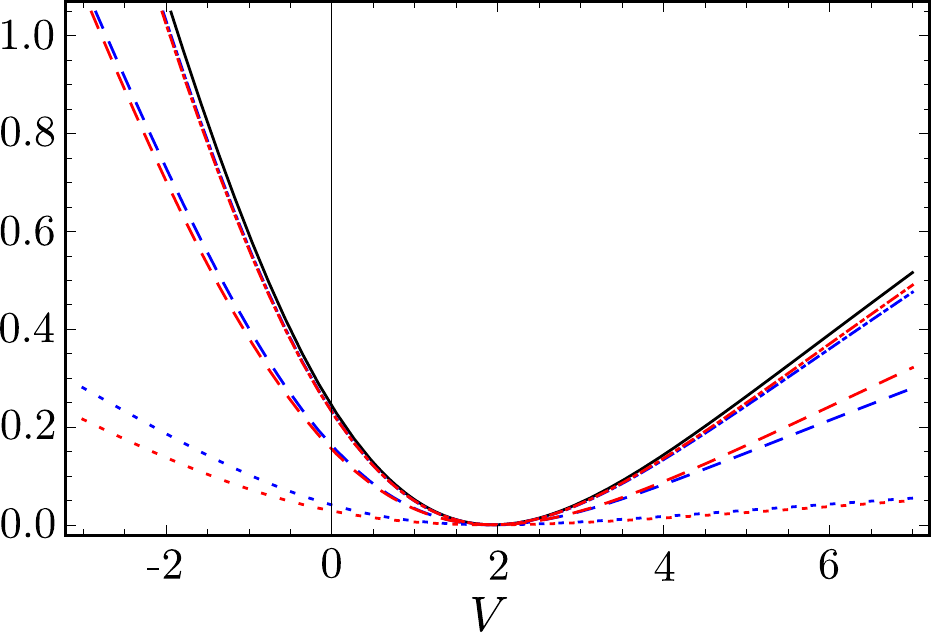}
\caption{\label{fig infinite fast feedback} Plot of the entropy production rates 
$\dot{\tilde S}_\bb{i}$ [Eq. (\ref{eq 2nd law effective effective}), black solid line] as well as  
$\dot S_{\bb i}$ [Eq. (\ref{eq 2nd law effective}), blue lines] and $\Delta_\bb{i} S_\text{tape}/\tau$ 
[Eq. (\ref{eq 2nd law tape model}), red lines] for $\gamma=0.1, \tau=10$ (dotted lines), 
$\gamma=1, \tau=1$ (dashed lines) and $\gamma=10, \tau=0.1$ (dash-dotted lines) over the bias voltage 
$V$. We chose $\delta^- = 0.75$. }
\end{figure}

As a first crosscheck of the validity of Eq. (\ref{eq effective effective ME}), any computer algebra program can confirm 
\begin{equation}
 \lim_{\tau\rightarrow0} \frac{\Delta N^L}{\tau} = \lim_{\gamma\rightarrow\infty} I^L = \tilde I^L
\end{equation}
where $\Delta N^L$ was defined in Eq. (\ref{eq particle number L}). 
Together with Eq. (\ref{eq equivalence effective and tape}) it should be also immediately clear that 
\begin{equation}
 \lim_{\tau\rightarrow0} \frac{(\Delta_{\bb i}S_\text{tape})}{\tau} = \lim_{\gamma\rightarrow\infty} \dot S_{\bb i} = \dot{\tilde S}_{\bb i}.
\end{equation}
By constrast, the mutual information (\ref{eq mutual information}) becomes constant (and non-zero) 
for $\tau\rightarrow0$ such that its rate diverges: $\lim_{\tau\rightarrow0} I(S;M')/\tau = \infty$. 
Thus, the second law involving discrete feedback, Eq. (\ref{eq 2nd law feedback}), no longer yields a useful 
bound. This does not contradict the claim that the mutual information can give 
a tighter bound for the error-free case, see Sec. \ref{sec comparison two second laws}, 
because for $\delta^- \rightarrow 1$ the feedback affinity $\C A$ also becomes unbounded. Results are plotted 
in Fig. \ref{fig infinite fast feedback}. 

\section{Discussion}
\label{sec conclusions}

We now briefly summarize and discuss our results. 
We divide the discussion into three main parts, which can be related to Secs. 
\ref{sec physical Implementation of the model} and \ref{sec thermodynamics} (i), 
Sec. \ref{sec measurement based feedback scheme} (ii) and Sec. \ref{sec effective master equation} (iii). 

(i) We have constructed a solvable physical model able to rectify thermal fluctuations to transport 
electrons against a bias (i.e., to charge a battery) by writing information on a tape without energy expenditure. 
This `information-electric device' is thermodynamically equivalent to the model in Ref. 
\cite{MandalJarzynskiPNAS2012}. The physical treatment of the problem allows us to clearly state the assumptions 
needed to obtain this ideal equivalence. First, to have an unchanged first law of thermodynamics, we need to assume 
a frictionless gliding tape of bits and degenerate spin eigenstates in the spin valve and on the tape. Second, to 
keep the discussion at a moderate level, we have neglected various experimentally important imperfections, which would 
reduce the overall efficiency of our device. Namely, we have assumed perfectly opposite polarized leads, neglected 
phonon interactions and assumed a fully unitary and error-free swap gate. 
Although we used the physical picture of a 
spin valve interacting via a Heisenberg Hamiltonian with the bits, it is worth pointing out that this could be 
also realized with other models: aside from the Brownian ratchet analogue, one could, for instance, use 
double quantum dots where the left and right eigenstates form a pseudospin giving rise to the same system dynamics, 
and -- as pointed out at the end of Sec. \ref{sec system bit interaction} -- one does not necessarily need to 
implement a tape of bits although this picture might still be helpful for thermodynamic considerations. 
Finally, it is interesting to note the similarity of these information driven devices with the theory of the 
micromaser \cite{ScullyZubairyBook} in which a stream of excited two-level atoms (like a tape of bits) is injected 
into a high quality cavity building up a photon field. Indeed, the device proposed by Scully in Ref. \cite{ScullyPRL2001} 
can be regarded as a quantum optical version of an information driven engine, where the entropy of the motional degrees of 
freedom of the two-level atoms is increased while energy is continuously extracted from the population of the excited 
states. 

(ii) We have formulated a discrete measurement and feedback model with identical system dynamics but a conceptually 
different second law of thermodynamics. Indeed, there are multiple second laws 
due to the fact that a dynamically irrelevant parameter ($\delta_0$) entered the mutual information. This 
additional parameter might be more rigorously exploited to give a stronger bound on the extractable work by 
minimizing the mutual information with respect to $\delta_0$, although its precise value $\delta_0^\text{min}$ 
might not be physically intuitive [Eq. (\ref{eq max min delta0})]. Furthermore, although we used identical 
`information resources' (namely a two-state memory equivalent to a bit on the tape), the second law for discrete 
feedback can provide a smaller, equal or larger bound than the second law involving the tape of bits, hence 
it remains a challenge to treat the different forms of information processing in one unified framework. 

(iii) We have derived an effective ME governing the system evolution for Poisson distributed events where an 
`event' can refer either to an interaction with a bit or a measurement and feedback step. The ME is the same for both 
approaches and we believe that its effective entropy production coincides with the entropy production 
of the tape model. \emph{A priori} this is not obvious, because we have seen that the mutual information can coincide 
with the change in Shannon entropy of the bits [see Sec. \ref{sec measurement based feedback scheme} or point (ii)]. 
However, we have shown that the \emph{a posteriori} postulated information current equals the rate of entropy 
change of an idealized tape of bits, see Eq. (\ref{eq equivalence effective and tape}), thus giving the 
phenomenologically introduced information current a precise physical and information theoretic basis. Furthermore, 
in the limit of infinite fast feedback we have demonstrated that the effective entropy production coincides with 
the entropy production of the tape model whereas the rate of mutual information diverges. It is worth mentioning two 
more things: first of all, the information current (\ref{eq equivalence effective and tape}) has the standard form often 
encountered in thermodynamics of a current times an affinity. Thus, the finding in Ref. \cite{BaratoSeifertPRL2014} -- 
that the entropy production involving the Shannon entropy difference of the information reservoir has an \emph{extra} 
term different from the usual current times affinity -- is in general not true. Second, we want to mention that under 
certain assumptions it is indeed possible to define a rate of mutual information as was done, for instance, in Ref. 
\cite{HorowitzSagawaParrondoPRL2013}. Note, however, that this has not the standard form of a current times an affinity 
and is always positive and thus, makes it again impossible to address question of information erasure or related issues. 

\section*{Acknowledgments}

PS thanks Sebastian Deffner and Zhiyue Lu for many stimulating discussions. Financial support by the DFG 
(SCHA 1646/2-1, SFB 910, and GRK 1558) is gratefully acknowledged. 
CJ gratefully acknowledges financial support from the National Science Foundation (USA) under grant DMR-1206971. 


\begin{thebibliography}{33}
 \bibitem{Maxwell1871} J. C. Maxwell, \emph{Theory of Heat} (1871). 
 \bibitem{Landauer1961} R. Landauer, IBM J. Res. Dev. \bb{5}, 183 (1961).
 \bibitem{Bennett1982} C. H. Bennett, Int. J. Theor. Phys. \bb{21}, 905-940 (1982).
 \bibitem{Szilard1929} L. Szilard, Zeitschrift f\"ur Physik \bb{53}, 840-856 (1929).
 \bibitem{Smulochowski1912} M. Smulochowski, Zeitschrift f\"ur Physik \bb{13}, 1069-1080 (1912).
 \bibitem{FeynmanLectures1963} R. P. Feynman, R. Leighton, and M. Sands, \emph{The Feynman Lectures on Physics} (Addison-Wesley, Reading, 1963).
 \bibitem{NortonEntropy2013} J. D. Norton, Entropy \bb{15}, 4432-4483 (2013).
 \bibitem{SkordosZurek1992} P. A. Skordos and W. H. Zurek, Am. J. Phys. \bb{60}, 876 (1992).
 \bibitem{StrasbergSchallerPRL2013} P. Strasberg, G. Schaller, T. Brandes, and M. Esposito, Phys. Rev. Lett. \bb{110}, 040601 (2013).
 \bibitem{MandalJarzynskiPNAS2012} D. Mandal and C. Jarzynski, Proc. Natl. Acad. Sci. \bb{109}, 11641 (2012).
 \bibitem{BaratoSeifertEPL2013} A. C. Barato and U. Seifert, Europhys. Lett. \bb{101}, 60001 (2013).
 \bibitem{HorowitzSagawaParrondoPRL2013} J. M. Horowitz, T. Sagawa, and J. M. R. Parrondo, Phys. Rev. Lett. \bb{111}, 010602 (2013).
 \bibitem{MandalQuanPRL2013} D. Mandal, H. T. Quan, and C. Jarzynski, Phys. Rev. Lett. \bb{111}, 030602 (2013).
 \bibitem{DeffnerJarzynskiPRX2013} S. Deffner and C. Jarzynski, Phys. Rev. X \bb{3}, 041003 (2013).
 \bibitem{DeffnerPRE2013} S. Deffner, Phys. Rev. E \bb{88}, 062128 (2013).
 \bibitem{HoppenauEngelEPL2014} J. Hoppenau and A. Engel, Euro. Phys. Lett. \bb{105}, 50002 (2014).
 \bibitem{DattaNotes2007} S. Datta, \emph{Nanodevices and Maxwell's Demon}, Lecture Notes, arXiv: 0704.1623. 
 \bibitem{LuMandalJarzynskiPT2014} Z. Lu, D. Mandal, and C. Jarzynski, Physics Today \bb{67}(8), 60 (2014).
 \bibitem{WisemanMilburnBook} H. M. Wiseman and G. J. Milburn, \emph{Quantum Measurement and Control} (Cambridge University Press, Cambridge, 2010).
 \bibitem{SchallerEtAlPRB2011} G. Schaller, C. Emary, G. Kiesslich, and T. Brandes, Phys. Rev. B \bb{84}, 085418 (2011).
 \bibitem{EspositoSchallerEPL2012} M. Esposito and G. Schaller, Euro. Phys. Lett. \bb{99}, 30003 (2012).
 \bibitem{StrasbergSchallerPRE2013} P. Strasberg, G. Schaller, T. Brandes, and M. Esposito, Phys. Rev. E \bb{88}, 062107 (2013).
 \bibitem{MunakataRosinbergJSM2012} T. Munakata and M. L. Rosinberg, J. Stat. Mech., P05010 (2012). 
 \bibitem{MunakataRosinbergJSM2013} T. Munakata and M. L. Rosinberg, J. Stat. Mech., P06014 (2013).
 \bibitem{TasakiArXiv2013} H. Tasaki, arXiv: 1308.3776. 
 \bibitem{BaratoSeifertPRL2014}  A. C. Barato and U. Seifert, Phys. Rev. Lett. \bb{112}, 090601 (2014).
 \bibitem{HorowitzEspositoPRX2014} J. M. Horowitz and M. Esposito, Phys. Rev. X \bb{4}, 031015 (2014).
 \bibitem{ShiraishiSagawaArXiv2014} N. Shiraishi and T. Sagawa, arXiv: 1403.4018.
 \bibitem{BauerBaratoSeifertArXiv2014} M. Bauer, A. C. Barato, and U. Seifert, J. Stat. Mech., P09010 (2014). 
 \bibitem{BoganiWernsdorferNatMat2008} L. Bogani and W. Wernsdorfer, Nat. Mater. \bb{7}, 179 (2008).
 \bibitem{UrdampilettaEtAlNatMat2011}  M. Urdampilleta, S. Klyatskaya, J.-P. Cleuziou, M. Ruben, and W. Wernsdorfer, Nat. Mater. \bb{10}, 502 (2011).
 \bibitem{BraunKoenigMartinekPRB2004} M. Braun, J. K\"onig, and J. Martinek, Phys. Rev. B \bb{70}, 195345 (2004).
 \bibitem{FuldeBook} P. Fulde, \emph{Electron Correlations in Molecules and Solids} (Springer-Verlag, 3rd. Edition, Berlin, 1995).
 \bibitem{LossDiVincenzoPRA1998} D. Loss and D. P. DiVincenzo, Phys. Rev. B \bb{57}, 120 (1998).
 \bibitem{PettaEtAlScience2005} J. R. Petta \emph{et al.}, Science \bb{309}, 2180 (2005).
 \bibitem{ToyabeEtAlNatPhys2010} S. Toyabe, T. Sagawa, M. Ueda, E. Muneyuki, and M. Sano, Nat. Phys. \bb{6}, 988 (2010). 
 \bibitem{CaoFeitoPRE2009} F. J. Cao and M. Feito, Phys. Rev. E \bb{79}, 041118 (2009).
 \bibitem{DeffnerLutzArXiv2012} S. Deffner and E. Lutz, arXiv: 1201.3888.
 \bibitem{CoverThomasBook} T. M. Cover and A. J. Thomas, \emph{Elements of Information Theory} (Wiley-Interscience, Hoboken, NJ, 2006).
 \bibitem{KanazawaSagawaPRE2013} K. Kanazawa, T. Sagawa, and H. Hayakawa, Phys. Rev. E \bb{87}, 052124 (2013).
 \bibitem{DianaEspositoJSM2014} G. Diana and M. Esposito, J. Stat. Mech., P04010 (2014). 
 \bibitem{HartichBaratoSeifertJSM2014} D. Hartich, A. C. Barato, and U. Seifert, J. Stat. Mech., P02016 (2014).
 \bibitem{ScullyZubairyBook} M. O. Scully and M. Suhail Zubairy, \emph{Quantum Optics} (Cambridge University Press, Cambridge, 1997).
 \bibitem{ScullyPRL2001} M. O. Scully, Phys. Rev. Lett. \bb{87}, 220601 (2001). 
 \bibitem{KosloffEnt13} R. Kosloff, Entropy \bb{15}, 2100-2128 (2013).
\end{thebibliography}

\appendix

\section{Proof of \texorpdfstring{$\Delta W_\text{pull}=0$}{}}
\label{sec app Proof Delta W zero}

We denote the combined system and bit Hamiltonian during one swap operation by 
$H_{SB}(t) \equiv H_S + H_B + V(t)$ where $t\in[t_-,t_+]$ with $t_+=t_-+\delta t$ and $V$ 
denotes the system-bit coupling generated by the Heisenberg interaction fulfilling
$V(t_-) = 0 = V(t_+)$. The change in system and bit energy is  given by 
\begin{equation}
 \begin{split}
  \frac{d}{dt} E(t)	&=	\frac{d}{dt}\mbox{tr}_{SB}[H_{SB}(t)\rho_{SB}(t)]	\\
			&=	\mbox{tr}_{SB}[\dot H_{SB}(t)\rho_{SB}(t) + H_{SB}(t)\dot\rho_{SB}(t)]	\\
			&\equiv	\dot W_\text{pull}(t) + \dot Q(t).
 \end{split}
\end{equation}
In the general theory of open quantum systems the first term on the right hand side is interpreted as the work and the 
second as the heat flow \cite{KosloffEnt13}. Because we assume the system to change unitarily we have 
\begin{equation}
 \begin{split}
  \dot Q	&=	\mbox{tr}_{S+B}[H_{S+B}(t)\dot\rho_{S+B}(t)]	\\
		&=	-i\mbox{tr}\left\{H_{S+B}(t)[H_{S+B}(t),\rho(t)]\right\} = 0.
 \end{split}
\end{equation}
Thus, after integrating over one swap operation from $t_-$ to $t_+$, we have 
$E(t_+) - E(t_-) = \Delta W_\text{pull}$ and hence, 
\begin{equation}
 \begin{split}
  \Delta W_\text{pull}	=&~	\mbox{tr}_{SB}[H_{SB}(t_+)\rho_{SB}(t_+) - H_{SB}(t_-)\rho_{SB}(t_-)]	\\
			=&~	\mbox{tr}_{SB}[\{H_S(t_+)+H_B(t_+)\}\rho_{SB}(t_+)]	\\
			&-	\mbox{tr}_{SB}[\{H_S(t_-)+H_B(t_-)\}\rho_{SB}(t_-)]	\\
			=&~	0
 \end{split}
\end{equation}
because $V(t_-) = 0 = V(t_+)$. Note that these results remains true even if we introduce a certain asymmetry $\Delta$ in the 
energy of spin up and down for both system and bit Hamiltonian 
[i.e., $H_{S/B} = \epsilon_{s/b} |\ua\rangle\langle\ua| + (\epsilon_{s/b} + \Delta)|\da\rangle\langle\da|$], but this would 
eventually affect the first law of thermodynamics for the system (or bit) alone. 

\section{Proof of \texorpdfstring{$\Delta_{\bb i} S_\text{tape} \ge 0$}{}}
\label{sec app Proof 2nd law tape model}

Deducing the second law for our model is not as clear as one might expect, especially because our dynamics 
are mixed between a unitary evolution and a subsequent dissipative (ME like) evolution. Arguments making use 
of relative entropy (or the Kullback Leibler divergence) as in Ref. \cite{MandalQuanPRL2013} or arguments 
relying on an integral fluctuation theorem do not apply as easily to our setup. 
We will thus explicitly prove the second law in the same way as done by Mandal and Jarzynski 
\cite{MandalJarzynskiPNAS2012}. The proof proceeds in two steps: first, we will show Eq. 
(\ref{eq 2nd law tape model}) for the stationary case $\tau\rightarrow\infty$ 
and we will then use this result to prove it for any finite $\tau$. 

Note that for our proof we only have to focus on the regions 1 and 2 in Fig. \ref{fig phase diagram}. 
In region 3 we have $V\Delta N^L>0$ \emph{and} $\Delta H_B>0$ such that Eq. (\ref{eq 2nd law tape model}) 
is trivially fulfilled and regions $1^\prime$, $2^\prime$ and $3^\prime$ follow from regions 1,2 and 3 by symmetry. 

The advantage of the long time limit $\tau\rightarrow\infty$ is that many expressions become quite simple. 
In particular, we have 
\begin{align}
 \Delta N^L(\infty)	&=	-\frac{V}{2} \frac{\delta^- + 1 + e^V (\delta^- - 1)}{1+e^{V/2}+e^V},	\\
 \delta^+(\infty)	&=	\frac{e^V-1+e^{V/2}\delta^-}{1+e^{V/2}+e^V}.
\end{align}
where $f(\infty) \equiv \lim_{\tau\rightarrow\infty} f(\tau)$ for any function $f$ of $\tau$ 
(if the limit exists). 
To simplify the subsequent algebra it is convenient to rescale the variables. First, 
we introduce $x\equiv e^{V/2}$ such that $x\in[1,\infty)$ for $V$ in region 1,2. Next, we parametrize 
$\delta^-$ by another parameter $h$ through 
\begin{equation}
 \delta^- = \tanh\frac{V}{2} + \frac{h}{e^V+1} = \frac{x^2-1}{x^2+1} + \frac{h}{x^2+1}.
\end{equation}
For $h=0$ we are on the line $\delta^+ = \delta^-$, which separates region 1 and 2, see Fig. \ref{fig phase diagram}. 
Furthermore, one can show that for $\tau\rightarrow\infty$ the line $\delta^+ = -\delta^-$ is given by 
$\delta^- = -\tanh\frac{V}{4}$ such that $h$ is bounded in region 1 and 2 by 
\begin{equation}\label{eq bounds h}
 \begin{split}
  \text{region 1: }	&	-2\frac{x^3-1}{x+1} \le h \le 0,	\\
  \text{region 2: }	&	0 \le h \le 2.
 \end{split}
\end{equation}
Next, we have a look at the derivative of $\Delta_{\bb i} S_\text{tape}$ with respect to $h$, which reads: 

\begin{widetext}

\begin{equation}
 \Delta_{\bb i} S'_\text{tape} \equiv \frac{\partial}{\partial h}\Delta_{\bb i} S_\text{tape} = \frac{\left(x^2+x+1\right) \mbox{arctanh}\left(\frac{h-2}{x^2+1}+1\right)-x \mbox{arccoth}\left(\frac{\left(x^2+1\right)\left(x^2+x+1\right)}{(h-1)x+x^4+x^3-1}\right)-\left(x^2+1\right) \ln (x)}{x^2+x+1}
\end{equation}
and we note that $\Delta_{\bb i} S'_\text{tape}(h=0) = 0$, which can be deduced by using the identities 
\begin{equation}
 \mbox{arctanh}(z) = \frac{1}{2}\ln\frac{1+z}{1-z}, ~~~ \mbox{arccoth}(z) = \frac{1}{2}\ln\frac{z+1}{z-1}.
\end{equation}
We are done with our proof if we can show that $\Delta_{\bb i} S'_\text{tape} \le 0$ for region 1 and 
$\Delta_{\bb i} S'_\text{tape} \ge 0 $ for region 2 because this implies $\Delta S_{\bb i} \ge 0$ for region 1 and 2. 
To deduce this we have a look at the second derivative, which becomes 
\begin{equation}
 \Delta_{\bb i} S''_\text{tape} \equiv \frac{\partial^2}{\partial h^2}\Delta_{\bb i} S_\text{tape} = \frac{2 (x+1)(x^2+1)^2 [h(1-x) + 2x(x+1)]}{(h-2) (h+2x^2) [h + 2x(x^2+x+1)] [hx - 2(x^2+x+1)]}.
\end{equation}
\end{widetext}
If we can show that $\Delta_{\bb i} S''_\text{tape} \ge 0$ for all $h$ in region 1 and 2 and for all $x\ge 1$, we are done 
with the proof because this would imply that $\Delta_{\bb i} S'_\text{tape}$ is a monotonically increasing function and 
since $\Delta_{\bb i} S'_\text{tape}(h=0) = 0$ we have $\Delta_{\bb i} S'_\text{tape} \le 0$ in region 1 and 
$\Delta_{\bb i} S'_\text{tape} \ge 0$ 
in region 2 as desired. To prove $\Delta_{\bb i} S''_\text{tape} \ge 0$ it suffices to look at the sign of all the factors 
of $\Delta_{\bb i} S''_\text{tape}$ and to show that their overall sign is positive. The estimation of the factors can be 
straightforwardly done by using the bounds for $h$, Eq. (\ref{eq bounds h}), which we will not do here. Thus, 
we have proven that Eq. (\ref{eq 2nd law tape model}) is true for $\tau\rightarrow\infty$. 

Let us now turn to the case of finite $\tau$. To begin, we have confirmed numerically that 
\begin{equation}\label{eq help useful identity 3}
 \Delta N^L(\tau) = \eta \Delta N^L(\infty) ~~~ \text{with} ~~~ \eta\in[0,1],
\end{equation}
i.e., the number of tunneled particles becomes maximized (in absolute value) for $\tau\rightarrow\infty$. 
Next, for $\tau\rightarrow\infty$ Eq. (\ref{eq help useful identity 2}) implies 
$\Delta N^L(\infty) = (\delta^+(\infty) - \delta^-)/2$ and again using Eq. (\ref{eq help useful identity 2}) 
for finite $\tau$ together with Eq. (\ref{eq help useful identity 3}) gives 
\begin{equation}
 \begin{split}
  \delta^+(\tau)	&=	\delta^- + 2\eta\frac{\delta^+(\infty) - \delta^-}{2}	\\
			&=	(1-\eta)\delta^- + \eta\delta^+(\infty).
 \end{split}
\end{equation}
Using that the entropy is a concave function (i.e., $\partial_\delta^2 H[\delta] < 0$) yields 
\begin{equation}
 \begin{split}
  H_B[\delta^+(\tau)]	&\ge	(1-\eta)H_B[\delta^-] + \eta H_B[\delta^+(\infty)]	\\
			&=	H_B[\delta^-] + \eta\left\{H_B[\delta^+(\infty)] - H_B[\delta^-]\right\}	\\
			&=	H_B[\delta^-] + \eta\Delta H_B(\infty),
 \end{split}
\end{equation}
which, together with the second law in the long time limit, implies 
\begin{equation}
 \begin{split}
  H_B[\delta^+(\tau)]	&\ge	H_B[\delta^-] - \eta V\Delta N^L(\infty)	\\
			&=	H_B[\delta^-] - V\Delta N^L(\tau)
 \end{split}
\end{equation}
and, after rearrangement, we finally obtain our desired result for all $\tau$ 
\begin{equation}
 \Delta_{\bb i} S_\text{tape}(\tau) = V\Delta N^L(\tau) + \Delta H_B(\tau) \ge 0. ~~~ \text{Q.E.D.}
\end{equation}

\section{Maximum and Minimum of the Mutual Information}
\label{sec app max min mutual info}

The first derivative of the mutual information with respect to $\delta_0$ is given by 
\begin{equation}
 \begin{split}
  I'	\equiv&~	\frac{\partial I(S;M')}{\partial\delta_0}	\\
	=&~		(1 - p_\ua^- - p_\da^-)	\\
	&\times		\left\{\mbox{arctanh}(\delta_0) - \mbox{arctanh}[h(\delta_0,\delta^-)]\right\}
 \end{split}
\end{equation}
with $h(\delta_0,\delta^-) \equiv \delta_0 - \delta_0(p_\ua^- + p_\da^-) + \delta^-(p_\ua^- - p_\da^-)$. Clearly, 
$I' = 0$ if $\delta_0 = \delta_0^\text{min}$ with $\delta_0^\text{min}$ from Eq. (\ref{eq max min delta0}) and 
it can easily be checked that the second derivative is positive at this point, hence $\delta_0^\text{min}$ 
truly minimizes the mutual information. We also recognize that for $\delta_0>\delta_0^\text{min}$ we have 
$I'>0$ and for $\delta_0<\delta_0^\text{min}$ we have $I'<0$ for every $\delta^-$. Thus, the mutual information 
has its local maxima at $\delta_0=\pm1$. From Eq. (\ref{eq delta M feedback}) we obtain 
\begin{align}
 \delta_M(\delta_0=+1)	&=	p_0^- + \delta^-(p_\ua^- - p_\da^-),	\\
 \delta_M(\delta_0=-1)	&=	-p_0^- + \delta^-(p_\ua^- - p_\da^-).
\end{align}
We now claim that 
$p_\ua^-\ge p_\da^-$ in the Maxwell demon region 2 of Fig. \ref{fig phase diagram}. Intuitively this should be 
clear because a positive bias $V>0$ favors the spin up state [see Eq. (\ref{eq steady state spin valve})] 
and the excess of spin up states on the tape will also preferably flip spin down states to spin up states 
in the system. Mathematically, however, we could only confirm this numerically. Accepting $p_\ua^-\ge p_\da^-$ 
we easily see that $|\delta_M(\delta_0=+1)|\ge|\delta_M(\delta_0=-1)|$, which implies that 
$I(\delta_0=-1) = H[\delta_M(\delta_0=-1)] - p_1^- H[\delta^-]$ is larger than 
$I(\delta_0=+1) = H[\delta_M(\delta_0=+1)] - p_1^- H[\delta^-]$.

\section{Derivation of the effective Master Equation}
\label{sec app derivation eff ME}

We divide the evolution of the system density matrix into pieces where there is either no interaction with the bit 
(only the dissipative dynamics $\C W$ of the spin valve) or there is a sudden interaction with a bit 
(and no dissipation) due to the swap operation. We introduce the notation 
$\C J\rho_S \equiv \mbox{tr}_B[U(\rho_S\otimes\rho_B)U^\dagger]$ with the same $U$ from Eq. 
(\ref{eq system time evolution 2}) and call $\C J$ a `jump operator'. Because we are still working in the basis 
$\rho_S = (p_0,p_\ua,p_\da)^T$ and neglect any off-diagonal elements of the density matrix, we can represent $\C J$ 
by a matrix of the form 
\begin{equation}
 \C J = 
 \left(\begin{array}{ccc}
        1	&	0		&	0		\\
        0	&	p_{b=\ua}^-	&	p_{b=\ua}^-	\\
        0	&	p_{b=\da}^-	&	p_{b=\da}^-	\\
       \end{array}\right) = 
 \left(\begin{array}{ccc}
        1	&	0			&	0			\\
        0	&	\frac{1+\delta^-}{2}	&	\frac{1+\delta^-}{2}	\\
        0	&	\frac{1-\delta^-}{2}	&	\frac{1-\delta^-}{2}	\\
       \end{array}\right).
\end{equation}
Note that the jump operator for the situation involving measurement and feedback [together with the choice 
(\ref{eq meas error})] is exactly the same, hence the ME is also the same for both cases. 

The density matrix $\rho_S^{(n)}(t)$ describing the state of the system after $n$ jumps have occurred up to time $t$ is 
then given by 
\begin{widetext}

\begin{equation}
 \begin{split}
  \rho_S^{(n)}(t)	&=	\int_{t_0}^t dt_n\int_{t_0}^{t_n} dt_{n-1} \dots \int_{t_0}^{t_2} dt_1 \gamma e^{-\gamma(t-t_n)} e^{\C W(t-t_n)} \C J \dots \C J \gamma e^{-\gamma(t_1-t_0)} e^{\C W(t_1-t_0)} \rho_S^{(0)}(t_0)	\\
			&=	\gamma^n e^{-\gamma(t-t_0)} \int_{t_0}^t dt_n\int_{t_0}^{t_n} dt_{n-1} \dots \int_{t_0}^{t_2} dt_1 e^{\C W(t-t_n)} \C J \dots \C J e^{\C W(t_1-t_0)} \rho_S^{(0)}(t_0)
 \end{split}
\end{equation}
\end{widetext}
and the average state of the system can be recovered via $\rho_S(t) = \sum_{n=0}^\infty \rho_S^{(n)}(t)$. 

Next, we take the time derivative of $\rho_S^{(n)}$, which yields three terms: 
\begin{equation}
 \frac{\partial}{\partial t} \rho_S^{(n)}(t) = -\gamma \rho_S^{(n)}(t) + \C W \rho_S^{(n)}(t) + \gamma\C J \rho_S^{(n-1)}(t).
\end{equation}
Averaging over $n$ gives finally the ME $\rho_S(t) = \C W_\text{eff} \rho_S(t)$ with the effective Liouvillian 
$\C W_\text{eff} = \C W_L + \C W_R + \C W_B$ where $\C W_{L,R}$ are given by Eq. (\ref{eq Liouvillian spin valve}) 
and $\C W_B \equiv \gamma(\C J - 1)$, which equals Eq. (\ref{eq effective ME}) of the main text. 

\section{Proof of Eq. (\ref{eq equivalence effective and tape})}
\label{sec app proof of Eq}

The change of the Shannon entropy of the bits over an infinitesimal small time step $dt$ is given by 
$dH_B(t) = H_B(t+dt)-H_B(t)$ with 
\begin{align}
 H_B(t+dt)	&=	-\sum_\sigma p_{b=\sigma}(t+dt)\ln p_{b=\sigma}(t+dt),	\\
 H_B(t)		&=	-\sum_\sigma p_{b=\sigma}(t)\ln p_{b=\sigma}(t)	
\end{align}
where $p_{b=\sigma}(t)$ denotes the probability to find the bit in state $\sigma\in\{\ua,\da\}$. Because at every 
time step the old bit is replaced by a new bit initialized with the probability distribution for the incoming tape, 
we must set $p_{b=\ua}(t) = (1+\delta^-)/2$ and $p_{b=\da}(t) = (1-\delta^-)/2$. The state of the outgoing bit 
is then given by 
\begin{equation}
 \begin{split}
  &	p_{b=\ua}(t+dt)	\\
  &=	p_{b=\ua}(t) + \gamma dt[p_{b=\da}(t)p_{s=\ua}(t) - p_{b=\ua}(t)p_{s=\da}(t)]	\\
  &=	p_{b=\ua}(t) + dt I^B
 \end{split}
\end{equation}
where we have explicitly denoted the state of the system at time $t$ with $p_{s=\sigma}(t)$ and $I^B$ is the current from 
Sec. \ref{sec Poisson distributed tape}. Similarly, $p_{b=\da}(t+dt) = p_{b=\da}(t) - dt I^B$. Using these relations 
and the expansion of the logarithm $\ln(1+x) = x + \C O(x^2)$, it is now a matter of straightforward algebra to show 
that 
\begin{equation}
 \begin{split}
  \frac{dH_B(t)}{dt}	&\equiv	\frac{H_B(t+dt)-H_B(t)}{dt}	\\
			&=	I^B \ln\frac{1-\delta^-}{1+\delta^-} + \C O(dt),
 \end{split}
\end{equation}
which proves Eq. (\ref{eq equivalence effective and tape}).

\end{document}